\documentclass[12pt,amsbsy]{article}
\usepackage{epsfig}

\newcommand{\beq}{\begin{equation}}
\newcommand{\eeq}{\end{equation}}

\newcommand{\beqn}{\begin{eqnarray}}
\newcommand{\eeqn}{\end{eqnarray}}

\newcommand{\al}{\mbox{${\alpha}$}}
\newcommand{\be}{\mbox{${\beta}$}}

\newcommand{\om}{\mbox{${\omega}$}}
\newcommand{\Om}{\mbox{${\Omega}$}}
\newcommand{\la}{\mbox{${\lambda}$}}
\newcommand{\La}{\mbox{${\Lambda}$}}
\newcommand{\si}{\mbox{${\sigma}$}}

\newcommand{\pa}{\mbox{${\partial}$}}

\begin{document}


\vspace{1cm}

\begin{center}
{\large \bf Does Cosmological Term Influence Gravitational Lensing?}
\end{center}

\begin{center}
I.B. Khriplovich\footnote{khriplovich@inp.nsk.su}, A.A.
Pomeransky\footnote{a.a.pomeransky@inp.nsk.su}
\end{center}
\begin{center}
Budker Institute of Nuclear Physics, 630090 Novosibirsk, Russia,\\
and Novosibirsk University
\end{center}

\bigskip

\begin{abstract}
We analyze the bending of light by galaxies or clusters of
galaxies in the presence of the $\La$ term. Going over to the
Friedmann-Robertson-Walker (FRW) coordinates, used in fact for the
description of actual observations, we demonstrate that the
cosmological constant does not influence practically the lensing
effect.
\end{abstract}

\bigskip

The effect of the cosmological constant $\La$ on the bending of
light by galaxies or clusters of galaxies was analyzed previously in
numerous papers. It was pointed out long ago in Ref. \cite{isl} that
in the Schwarzschild-de Sitter metric (derived in Ref.
\cite{kot})), which includes the $\La$-term, the latter does not
enter at all the
exact differential equation for the trajectory of a light ray. This
result was confirmed and elaborated upon in Refs. [3--6]. 
Recently, however, the assertion was made in Ref.
\cite{rin} that, though the above result is by itself correct, the
$\La$-term does contribute to the bending of light. This assertion,
in its turn, was reiterated and elaborated upon in Refs. [9--13].
Here we investigate the problem in the FRW
coordinates; they are the most relevant ones for the description of
actual observations.

\begin{figure}[htp]
\centering
\includegraphics[width=0.9\textwidth]{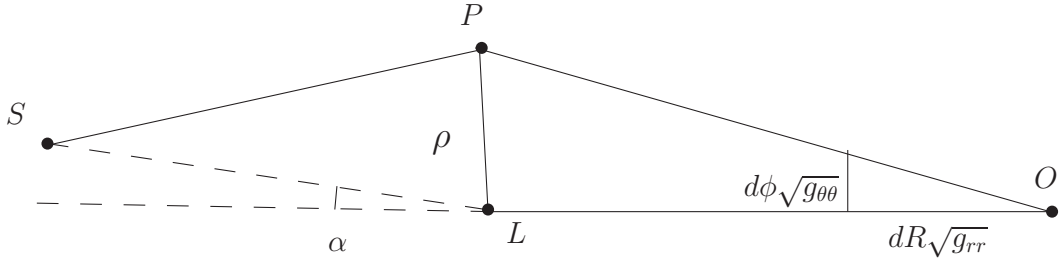}
\caption{Lensing geometry.}
\end{figure}

We start the analysis with the following simple statement. Let us
consider two light rays coming from the source $S$ to an observer
at the point $O$ (see Fig. 1).
One of the rays originates from the center $L$ of a gravitational
lens. Another one comes to the observer from the source, being
refracted by the lens; this ray crosses at the point
$P$ the plane passing through $L$ and
orthogonal to the axis lens -- observer, L -- O. Obviously, the
scalar product $g^{\mu\nu}k_{1\mu} k_{2\nu}$ is an
invariant. Let us consider this invariant in the locally inertial
frame where the observer is at rest. Here this scalar product of
the two null vectors, $k_1$ and $k_2$,
reduces (up to a factor of 2) to $\om^2 \theta^2$ where $\omega$ is
the light frequency and $\theta$ is the small angle between the rays.
Both $\omega$ and $\theta$ are directly measured by the observer at
rest in the locally inertial frame. We will work
with the square root of this expression (which is of course
an invariant also):
\beq\label{i1}
I = \om\,\theta.
\eeq

At first, let us consider this invariant in the dS coordinates.
We start with the Schwarz\-schild-de Sitter metric which describes
the space-time geometry outside a spherically symmetric lens~\cite{kot}:
\beq\label{ds}
ds^2 = (1 - \la^2 R^2-r_g/R)\, dT^2 - (1 - \la^2
R^2-r_g/R)^{-1}\,dR^2 - R^2 \,d\Om^2\,;
\eeq
here and below $r_g$ is the gravitational radius of the lens and
\[
\la^2=\La/3\,,\quad d\Om^2 = d\theta^2 + \sin^2 \theta\;
d\phi^2\,.
\]
The trajectory of a light ray in this space satisfies a simple
differential equation \cite{isl}: \beq\label{tra}
\frac{d\phi}{dR}=\frac{\rho}{R\sqrt{R^2-\rho^2+\rho^2r_g/R}}, \eeq
which can be easily derived from the Hamilton-Jacobi equation. Here
the integral of motion $\rho$ is the radial coordinate of the point
on the trajectory which is of the closest approach to the lens. In
the limit $R \gg \rho \gg r_g$, we are interested in, this equation
reduces to
\beq\label{tra1}
\frac{d\phi}{dR}=\frac{\rho}{R^2}\,.
\eeq
Integrating Eq. (\ref{tra}) between the source and the observer
one can express $\rho$ in terms of the radial positions of the
source and of the observer: \beq\label{rgr}
2r_g/\rho+\alpha=\rho/R_s+\rho/R_o, \eeq where small angle $\alpha$
is the difference between $\pi$ and the angle formed by the points
$S$, $L$ and $O$ (see Fig. 1).

Let us consider now a small triangle presented in Fig. 1 ($\theta
\ll 1$), with its vertex being the observation point. The angle
$\theta$, as seen from the vertex, is equal obviously to the ratio
of the two lengths, and transforms with eq. (\ref{tra1}) as follows:
\beq\label{an}
\theta =
\frac{d\phi\sqrt{|g_{\theta\theta}|}}{dR\sqrt{|g_{RR}|}} =
\frac{\rho}{R\sqrt{|g_{RR}|}} = \theta_0 \sqrt{1 - \la^2 R^2},
\eeq
where $\theta_0=\rho/R$. The factor $|g_{RR}|^{-1/2} = \sqrt{1 -
\la^2 R^2}$, is just the amendment, advocated in Ref. \cite{rin}, to
the common expression for the bending of light. Therefore, the
discussed invariant in the dS metric is
\beq\label{i3} I = \om_{\rm
dS}\,\theta_0 \sqrt{1 - \la^2 R^2}.
\eeq

We go over now to the FRW coordinates. This reference frame, where
the distances grow with time (though the coordinates by themselves
remain constant), is the most appropriate one for the description of observations,
and commonly used for this purpose.
Here the interval is
\beq\label{frw}
ds_{\rm FRW}^2 = dt^2 - a^2(t)(dr^2 + r^2 d\Om^2)\,,
\eeq
with
\beq\label{alt}
a(t)= e^{\lambda t}
\eeq
for the de Sitter Universe. It is convenient to choose the origin at
the point $L$ (as it has
been done above in the dS coordinates), and to start the count of
time $t$ at the moment of lensing.

To rewrite interval (\ref{i3}) in the FRW coordinates, we note
first of all that
\beq\label{c1}
R = a(t) r
\eeq
(which follows, for instance, from the comparison of the factors
at $d\Om^2$ in (\ref{ds}) and (\ref{frw})). Therefore,
\beq\label{c2}
\sqrt{1 - \la^2 R_o^2} = \sqrt{1 - \la^2 a^2(t_o) r_o^2};
\eeq
the subscript $o$ here and below means that the labelled symbols refer to the
point and moment of observation. Besides, angle $\theta_0$ in
equation (\ref{an}) can be written now as
\beq
\theta_0 = \,\frac{\rho}{a(t_o)\, r_o}\,,
\eeq
where $\rho$ is the impact parameter of the deflected photon.

Since $a(t)r$ is the distance of a point with a radial coordinate
$r$ from the origin (see (\ref{frw})), in FRW coordinates the
observer moves away from the origin (lens) with the velocity
\[
v = \dot{a}(t_o)\,r_o = \la\, a(t_o)\, r_o\,.
\]
Therefore, the frequency  $\om_{\rm \,FRW}$ received by the
observer is Doppler-shifted with respect to the static one
$\om_{\rm \,dS}$ as follows:
\[
\om_{\rm \,FRW} = \om_{\rm \, dS}\,\sqrt{\frac{1-\la a(t_o)
r_o}{1+\la a(t_o) r_o}}\,.
\]
In other words, we have to substitute
\[
\om_{\rm \,FRW}\,\sqrt{\frac{1+\la a(t_o) r_o}{1-\la a(t_o)
r_o}}
\]
for $\om_{\rm \, dS}$ in invariant (\ref{i3}). The product of the
last radical and that in formula (\ref{c2}) is equal to
\beq
1+\la a(t_o) r_o\,.
\eeq

At last, let us recall that the observation point $(t_o,\,r_o)$
belongs to the light cone with the top at the origin. This cone is
described by differential equation (see (\ref{frw}))
\beq\label{equ}
\frac{dt}{a(t)} = dr
\eeq
with the initial condition $r=0$ at $t=0$. With $a(t)= e^{\lambda
t}$ (see (\ref{alt})), the solution of equation (\ref{equ}) can be
easily rewritten as follows:
\beq\label{to}
a(t_o) = 1+\la a(t_o) r_o\,.
\eeq
Thus, in the FRW metric (\ref{frw}), used in practice for the
description of lensing effects, all factors in invariant $I$,
dependent on $\la$ and/or $a(t_o)$, cancel, so that this invariant
reduces to
\beq
I = \om_{\rm \,FRW}\,\frac{\rho}{r_o}\,.
\eeq

Then, the scale factor at the time of the light emission by
the source satisfies an equation analogous to (\ref{to}):
\beq\label{ts}
a(t_s) = 1-\la a(t_s) r_s\,
\eeq
(indeed, while the frequency received at the point $O$ is red-shifted 
with respect to the frequency $\om$ of the propagating signal, the 
frequency emitted at the point $S$ is red-shifted
with respect to $\om$).
It can be easily demonstrated with eqs. (\ref{c1}),
(\ref{to}), (\ref{ts}) that relation (\ref{rgr}) has in the FRW
coordinates quite analogous form:
\beq
2 r_g/\rho +\alpha =\rho/r_s+\rho/r_o\,.
\eeq
The angle $\alpha$, with its vertex at the origin, has obviously the same value
in the dS and FRW coordinates. Then, since 
\[
1/r_s+1/r_o = 1/R_s+1/R_o\,, 
\]
the parameter $\rho$ is also the same in the both frames discussed.

Thus, in the practically employed FRW metric all
``long-distance'' corrections caused by the cosmological constant,
i.e., those on the order of $\La r_o^2$ and higher (one could hardly
expect here corrections of the type $\la r_o \sim \sqrt{\La} r_o$,
nonanalitic in $\La$) to the effects of gravitational lensing cancel.
In particular,
the factor $\sqrt{1 - \la^2 R^2}$, pointed out in Ref. \cite{rin},
disappears here.

The absence of "long-distance" cosmological effects in gravitational lensing
is not at all surprising  and can be easily demonstrated to take place in a general case
of spatially-flat homogeneous and isotropic FRW Universe. The reason is that such space-times
are conformally-flat, and thus in the appropriate coordinates their metric is proportional
to the Minkowski one:
\[
ds_{\rm \,\,FRW}^2 = a^2(\eta) (d\eta^2 - dr^2 - r^2 d\Om^2).
\]
This interval can be obtained from Eq. (\ref{frw}) by changing the time coordinate according
to the relation $ d\eta=dt/a(t)$. The factor $a(\eta)$ does not influence the propagation of light rays,
since it cancels from the eikonal equation. Thus, the world lines of photons in this coordinate
frame are just the straight lines like in Minkowski space-time. We see here, that from the lensing
point of view there is nothing qualitatively special in the particular case of de Sitter space-time:
all cosmological effects are reduced to simply taking into account the scale factor at the moment of lensing
$a(t_L)$.

Let us note that some corrections on the order of $\la^2 \rho^2
\sim \La r_g r_o$ to the lensing effects may exist, as well as other
cosmological corrections in the general case of the FRW Universe.
However, such ``short-distance'' phenomena are perhaps too small to
be of practical interest.

To summarize, the effects of gravitational lensing are described
in the FRW coordinates with good accuracy by common formulas, both
in the presence or the absence of the cosmological term.

\end{document}